\def\spose#1{\hbox to 0pt{#1\hss}}
\def\lta{\mathrel{\spose{\lower 3pt\hbox{$\mathchar"218$}}
     \raise 2.0pt\hbox{$\mathchar"13C$}}}
\def\gta{\mathrel{\spose{\lower 3pt\hbox{$\mathchar"218$}}
     \raise 2.0pt\hbox{$\mathchar"13E$}}}
\def\ge{\mathrel{\spose{\lower 3pt\hbox{$-$}}
     \raise 2.0pt\hbox{$\mathchar"13E$}}}
\def\le{\mathrel{\spose{\lower 3pt\hbox{$-$}}
     \raise 2.0pt\hbox{$\mathchar"13C$}}}
\documentclass[aps,twocolumn]{revtex4}

\usepackage{graphics}

\begin{document}

\bibliographystyle{apsrev}

\title{Chaos and Gravitational Waves}
\author{Neil J. Cornish}
\affiliation{Department of Physics, Montana State University, Bozeman, MT 59717}

\begin{abstract}
The gravitational waveforms of a chaotic system will exhibit sensitive
dependence on initial conditions. The waveforms of nearby orbits decohere
on a timescale fixed by the largest Lyapunov exponent of the orbit. The loss
of coherence has important observational consequences for systems where the
Lyapunov timescale is short compared to the chirp timescale. Detectors 
that rely on matched filtering techniques will be unable to detect gravitational
waves from these systems.
\end{abstract}

\maketitle

%\narrowtext

The evolution of two compact objects under their mutual gravitational attraction
is a major unsolved problem in general relativity. A vast array of analytic and
numerical approximations have been used to attack the problem, but much remains
to be understood. Amongst the many outstanding questions is the degree to which
the dynamics may be chaotic\cite{us,maeda,janna,com} or effectively
chaotic\cite{scott,schutz}. For example, when one or more of the masses is spinning
it has been shown in the test-particle\cite{maeda} and post-Newtonian\cite{janna} limits
that certain classes of orbits are chaotic, at least when radiation reaction is
turned off\cite{com}.

Here we consider the physical and observational consequences wrought by chaotic
behaviour in the orbital dynamics. A key feature of chaotic systems is their sensitive
dependence on initial conditions - the so-called ``butterfly effect''. It
is intuitively obvious that sensitive dependence in the dynamics will be reflected in
the gravitational waveforms, but the precise connection has not been established until
now. We find that the waveforms of nearby orbits decohere on a timescale fixed by
the largest Lyapunov exponent. The observational consequences of this result depend on
two timescales: the chirp timescale $T_c$; and
the Lyapunov timescale $T_\lambda$. A third timescale of interest is the gravitational
wave period $T_w$. The wave period and chirp timescale for a system with
reduced mass $\mu$, total mass $M$ and orbital separation $R$ are given by
\[
T_w \simeq \pi M \left(\frac{R}{M}\right)^{3/2} \quad {\rm and} \quad 
T_c \simeq \frac{5}{256} \frac{M^2}{\mu}\left(\frac{R}{M}\right)^4 \, .
\]
In the limit that one of the masses is very much smaller than its companion we have $\mu \ll M$
and $T_c \gg T_w$. The Lyapunov timescale has to be calculated on a case-by-case basis, but
values as short as $T_\lambda \sim T_w$ are possible for very unstable orbits. Orbits whose
Lyapunov timescale is short compared to the chirp timescale will produces highly unpredictable
waveforms. The number of templates required to detect these waveforms is exponentially large,
making them impossible to detect\cite{janna,talk} with the Laser Interferometer
Gravitational Observatory (LIGO).

The connection between orbital instability and waveform decoherence is easily
established. It relies on the fact that the gravitational waveform $h_{\mu\nu}(t)$
can be expressed in terms of the phase space coordinates of the system, so the
divergence of nearby trajectories in phase space is reflected in the divergence
of nearby waveforms. The phase space coordinates we have in mind are the position
and momentum of the masses. These coordinates are well defined in the test particle,
post-Newtonian and post-Minkowski approximations, but their meaning is less clear
in the full theory. Indeed, the full evolution equations are a set of coupled,
non-linear partial differential equations, so the real issue is gravitational
turbulence, not chaos. We will avoid this complication by assuming that the
evolution equations can be approximated by some set of non-linear ordinary differential
equations for the phase space variables $X_{i}(t)$. Here $t$ is taken to be the time measured
by a distant observer, but the choice is unimportant. The evolution equations
can be written:
\begin{equation}
\frac{d X_{i}}{dt} = H_{i}(X_{j}) \, .
\end{equation}
Linearizing about a solution to this equation yields an equation for the perturbation:
\begin{equation}
\frac{d\,  \delta \! X_{i}(t)}{dt} = K_{ij}(t)\,  \delta \! X_{j}(t) \, ,
\end{equation}
where
\begin{equation}
K_{ij}(t) = \left. \frac{\partial H_{i}}{\partial X_{j}}\right|_{X_{i}(t)}
\end{equation}
is the infinitesimal evolution matrix. The solution to the linearized equations
of motion can be expressed in terms of the evolution matrix $L_{ij}(t)$:
\begin{equation}
\delta \! X_{i}(t) = L_{ij}(t)\, \delta\! X_{j}(0) \, .
\end{equation}
The eigenvalues and eigenvectors of $K_{ij}(t)$ contain information about the local
stability of the phase space trajectory. In a fixed coordinate system the
eigenvectors may vary with time. To avoid this, a non-constant
basis can be defined that stays aligned with the eigenvectors\cite{carl}. This is done by parallel
transporting the basis vectors along the trajectory $X_{i}(t)$. Barring degeneracies, the
infinitesimal evolution matrix is diagonal in the eigenvector basis:
$K_{ij}(t) = \delta_{ij} \lambda_{i}(t)$
(no summation on the $i$), where the $\lambda_{i}(t)$ are the eigenvalues of $K_{ij}(t)$.
The evolution matrix is also diagonal in the eigenvalue basis and has components
\begin{equation}
L_{ij}(t) = \delta_{ij} \exp\left(\int_0^t \lambda_{i}(t') dt'\right) \, .
\end{equation}
The principal Lyapunov exponent for the trajectory $X_{i}(t)$ is defined:
\begin{equation}\label{lyap}
\lambda = \lim_{t\rightarrow \infty}\frac{1}{2t}\ln\left(L_{ji}^{*}(t)L_{ij}(t)\right) \, .
\end{equation}
The multiplicative ergodic theorems establish the existence of this limit for a large
range of situations. Trajectories with positive Lyapunov exponents are
unstable against small perturbations. When the unstable orbits have non-zero measure
in phase space, the system is said to be chaotic. A positive principal Lyapunov exponent will
dominate the long term dynamics. Setting $\lambda_1$ to be the principle eigenvalue, the
asymptotic form of the evolution matrix can be written:
\begin{equation}\label{asym}
L_{ij}(t) \simeq e^{\lambda\, t} 
\left[ \begin{array}{cccc}
f(t) & 0 & 0 & \cdots \\
0    & 0 &  0 &\cdots \\
0    & 0 & 0 & \cdots \\
\vdots  & \vdots & \vdots  &  \ddots 
\end{array}
\right]
\end{equation}
where $\vert f(t) \vert=1$. If $X_1(t)$ is real, then $f(t)=1$.

A gravitational wave can be decomposed into two polarizations with amplitudes $h^+(t)$ and
$h^\times(t)$. A gravitational wave detector responds to the wave according to
\begin{equation}
s(t) = h^+(t) F_{+} + h^\times(t) F_\times \, ,
\end{equation}
where $F_+$ and $F_\times$ are the antenna patterns for each polarization. The
antenna patterns for LIGO and LISA can be found in Refs.\cite{flan} and \cite{paper1}
respectively. The gravitational wave amplitude $h(t)$ measured at the detector
can be expressed in terms of the phase space variables $X_{i}(t)$. Nearby orbits
have gravitational waveforms that differ by
\begin{eqnarray}\label{hpert}
\delta h(t) &=& \frac{\partial h(X_{m})}{\partial X_{i}}(t) \, \delta \! X_{i}(t)
\nonumber \\
&& \nonumber \\
&=& \frac{\partial h(X_{m})}{\partial X_{i}}(t)\,  L_{ij}(t)\, \delta\! X_{j}(0) \, .
\end{eqnarray}
Employing an eigenvector basis and using the asymptotic form (\ref{asym}) for the
evolution matrix yields
\begin{equation}
\delta h(t) \simeq e^{\lambda \, t} \, g(t) \, ,
\end{equation}
where
\begin{equation}
g(t) = \frac{\partial h(X_{m})}{\partial X_{1}}(t)\, f(t)\, \delta\! X_{1}(0)
\end{equation}
is an oscillatory factor with bounded amplitude. The divergence of the waveforms
can be better expressed in terms of their relative phase.
Writing the amplitude of the reference trajectory as $h(\Phi)$, and defining
$\Phi_0(t)$ to be the phase for which $h(\Phi_0)=0$, we find the relative phase
is given by
\begin{equation}
\delta \Phi(t) = \frac{\delta h(\Phi_0(t))}{h'(\Phi_0(t))} \, .
\end{equation}
The quantity
\begin{equation}\label{phase}
\lambda_h = \lim_{t \rightarrow \infty}\, \lim_{\delta\Phi(0)\rightarrow 0}
  \ln\left|\frac{\delta \Phi(t)}{\delta \Phi(0)}\right|
\end{equation}
is equal to the principal Lyapunov exponent of the trajectory. In other words, the waveforms
of nearby trajectories will decohere exponentially in time, on a timescale set by
$T_\lambda = \lambda^{-1}$.

A couple of comments are in order. Firstly, the limit
$t \rightarrow \infty$ used in equations (\ref{lyap}) and (\ref{phase}) will return
$\lambda=\lambda_h=0$ if we include the effects of radiation reaction. This reflects the
fact that the dynamics is not chaotic in a formal sense\cite{com}. The limit
$t \rightarrow \infty$ has to be replaced by the limit $t \rightarrow T$, where
$T_\lambda \ll T \ll T_c$. This is done by replacing the Lyapunov exponent by a
finite-time or local Lyapunov exponent\cite{ll}, and making a similar change to (\ref{phase}).
Clearly, the whole notion of waveform decoherence only makes sense if the
Lyapunov time scale $T_\lambda$ is very much shorter than the chirp timescale
$T_c$. It remains to be seen if this condition is satisfied for any realistic systems.
The second comment concerns the gauge non-invariance of the Lyapunov exponents and the choice of time
variable. Lyapunov exponents gained a bad reputation when they were
used to study the Bianchi IX dynamics. Different time slices yielded principal
Lyapunov exponents that were positive or zero, making it difficult to decide if
the dynamics was chaotic. However, this difficulty can be avoided by making relative,
rather than absolute comparisons. For example, the ratio $T_\lambda/T_c$ is the same
in any coordinate system, so the choice of time variable is irrelevant. The confusion
surrounding the Bianchi IX dynamics could have been avoided if the Lyapunov timescale
had been compared to the average time between bounces\cite{mix2}.

We can illustrate the connection between orbital instability and waveform decoherence
with a simple example. Consider the Lagrangian for a test particle in the Schwarzschild
spacetime (in units where $G=c=M=1$)
\begin{eqnarray}
{\cal L} &=& \frac{1}{2}\left( \frac{r-2}{r}\left(\frac{ d t}{d\lambda}\right)^2
+\frac{r}{r-2}\left(\frac{ d r}{d\lambda}\right)^2 \right. \nonumber \\
&& \quad \left. + r^2 \left(\frac{ d \theta}{d\lambda}\right)^2
+r^2\sin^2\theta \left(\frac{ d \phi}{d\lambda}\right)^2 \right).
\end{eqnarray}
The system is completely integrable (non-chaotic) as it has four generalized coordinates
$t,r,\theta,\phi$ and four constants of motion, the mass $\mu$, energy $E$, $z$-component
of angular momentum $L_z$ and total angular momentum squared $L^2$. Despite the lack of
chaos, the dynamics does admit isolated unstable orbits that serve to illustrate the connection between
orbital instability and waveform decoherence. Restricting
our attention to equatorial orbits with fixed energy $E$ allows us to make the replacement
\begin{equation}
\frac{d}{d \lambda} = \frac{r-2}{rE} \frac{d}{d t}
\end{equation}
in the equations of motion:
\begin{eqnarray}
\dot{p}_r &=& -\frac{ E^3 r}{(r-2)^3}-\frac{E p_r^2}{r(r-2)}+\frac{E p_\phi^2}{r^2(r-2)}
\nonumber \\
\dot{p}_\phi &=& 0 \nonumber \\
\dot{r} & =& E p_r \nonumber \\
\dot{\phi} & =& \frac{E p_\phi}{r(r-2)} \, .
\end{eqnarray}
Our reference trajectories are circular orbits with
\begin{eqnarray}
p_r &=& 0 \nonumber \\
p_\phi & = &  L_z = L \nonumber \\
r & = & r_\pm = \frac{1}{2}L(L\pm \sqrt{L^2 - 12})  \nonumber \\
\theta & = & \omega t \, .
\end{eqnarray}
Perturbing the equations of motion about the reference trajectory yields the
infinitesimal evolution matrix
\begin{equation}
K_{ij} = \left[ \begin{array}{cccc}
0 & \frac{2E L}{r^2(r-2)}& \frac{2E^3(r+1)}{(r-2)^4}-\frac{EL^2(3r-4)}{r^3(r-2)^2}& 0 \\
0    & 0 &  0 & 0 \\
E    & 0 & 0 & 0 \\
0 &  \frac{E}{r(r-2)} & -\frac{2EL(r-1)}{r^2(r-2)^2}&  0
\end{array}
\right] .
\end{equation}
The eigenvalues are:
\begin{eqnarray}
&& \lambda_1 = -\lambda_2 = \frac{\sqrt{2E^2r^3(r+1)-L^2(3r-4)(r-2)^2}}{r^2(r-2)^2}
\nonumber \\
&& \lambda_3 = \lambda_4 = 0 \, .
\end{eqnarray}
The well known innermost stable circular orbit at $r=6$ has $L=2\sqrt{3}$,
$E=\sqrt{8/9}$ and $\lambda_1=\lambda_2=0$. Circular orbits with larger values of $L$
come in two varieties, those with $r=r_+$ and those with $r=r_-$, e.g. setting
$E=1$ and $L=4$ yields orbits with $r_+=12$ and $r_-=4$. The outer orbit is stable,
$\lambda_1=-\lambda_2= i\sqrt{6}/108$, while the inner orbit is unstable,
$\lambda_1=-\lambda_2=1/2\sqrt{2}$. The unstable orbit at $r=4$ has eigenvectors
\begin{eqnarray}
{\bf e}_1 &=& \left(\frac{1}{3\sqrt{2}}, 0, \frac{2}{3}, -\frac{1}{\sqrt{2}}\right) \nonumber \\
{\bf e}_2 &=& \left(\frac{1}{3\sqrt{2}}, 0, -\frac{2}{3}, -\frac{1}{\sqrt{2}}\right) \nonumber \\
{\bf e}_3 &=& {\bf e}_4 = (1,0,0,0) \, .
\end{eqnarray}
Since the zero eigenvalues have degenerate eigenvectors, $K_{ij}$ can not be diagonalized,
but it is enough to put $K_{ij}$ into Jordan normal form with the transformation matrix
\begin{equation}
P_{ij} = \left[ \begin{array}{cccc}
\frac{7\sqrt{2}}{24} & -\frac{7\sqrt{2}}{24} & 0 & 0\\
 & & & \\
0    & 0 &  0 & 1 \\
 & & & \\
\frac{7}{6} & \frac{7}{6} & 0 & -2 \\
 & & & \\
-\frac{7\sqrt{2}}{8} & \frac{7\sqrt{2}}{8} & \frac{7}{8} & 0
\end{array}
\right] .
\end{equation}
In terms of the transformed phase space coordinates the infinitesimal
evolution matrix has components
\begin{equation}
K_{ij} = \left[ \begin{array}{ccccc}
\frac{1}{2\sqrt{2}} & 0 & 0 & \hspace*{0.06in} & 0 \\
 & & & \\
0 &-\frac{1}{2\sqrt{2}} & 0 &  & 0 \\
 & & & &\\
0  & 0 &  0 &  &1 \\
 & & & &\\
0    & 0 & 0 &  &0\\
\end{array}
\right] ,
\end{equation}
and the evolution matrix has components
\begin{equation}
L_{ij} = \left[ \begin{array}{ccccc}
e^{t/2\sqrt{2}}& 0 & 0 & \hspace*{0.06in} & 0 \\
 & & & &\\
0 &e^{-t/2\sqrt{2}} & 0 & & 0 \\
 & & & &\\
0  & 0 &  1 & & t \\
 & & & & \\
0    & 0 & 0 & & 1\\
\end{array}
\right] .
\end{equation}
Taking the limit defined in (\ref{lyap}) yields the principal Lyapunov exponent
$\lambda=1/2\sqrt{2}$. The Lyapunov timescale for this orbit, $T_\lambda=\lambda^{-1}=2\sqrt{2}$,
is actually {\em shorter} than the gravitational wave period $T_w=2\pi$.

Next we need an expression for the gravitational wave amplitude in terms of the
phase space variables. The correct approach would be to solve the Teukolsky
equation using the method described by Hughes\cite{scott}, but no analytic solutions
exist in the strong field limit. Instead we will use the quadrupole approximation,
knowing full well that it is being applied outside of its domain of validity. Since our
goal is to illustrate a general phenomenon - not to produce accurate templates for LIGO
data analysis - a qualitative description of the waveform's dependence on the phase space
variables will suffice. 

The gravitational wave amplitude can be written as $h_{\mu\nu}(t) = h^+(t)e^+_{\mu\nu}
+h^\times(t)e^\times_{\mu\nu}$, where the polarization tensors have non-zero components
\begin{eqnarray}
&& e^+_{xx}=1 \quad e^+_{yy}=-1   \nonumber \\
&& e^\times_{xy}=1 \quad e^\times_{yx}=1   \, .
\end{eqnarray}
A detector situated at $r=R$ on the $z$ axis will encounter a wave with plus polarization
given by
\begin{eqnarray}
h^+(t) &=& -\frac{2\mu E^2}{R}\left[\frac{2p_\phi p_r (r-3)}{(r-2)^2}\sin2\phi
+\left(\frac{E^2 r^2}{(r-2)^3} \right. \right.
\nonumber \\
&& \quad + \left.
\left. \frac{p_\phi^2(r+2)}{r(r-2)^2}
-\frac{p_r^2(r-3)}{r-2}\right)\cos2\phi 
\right] \, ,
\end{eqnarray}
and similarly for the cross polarization. Specializing to the
orbit with $E=1$, $L=4$ and $r=4$ we have
\begin{eqnarray}
h^+ &=& -\frac{8\mu}{R} \cos 2\phi \nonumber \\ 
&& \nonumber \\
\frac{\partial h^+}{\partial p_r} &=& -\frac{2\mu}{R} \sin 2\phi\nonumber \\ 
&& \nonumber \\
\frac{\partial h^+}{\partial p_\phi} &=& -\frac{3\mu}{R} \cos 2\phi\nonumber \\ 
&& \nonumber \\
\frac{\partial h^+}{\partial r} &=& \frac{17\mu}{2R} \cos 2\phi\nonumber \\ 
&& \nonumber \\
\frac{\partial h^+}{\partial \phi} &=& \frac{16\mu}{R} \sin 2\phi \, .
\end{eqnarray}
Putting everything together in equation (\ref{hpert}) yields
\begin{eqnarray}
\delta h^+(t) &\simeq& \frac{\mu}{84 R} e^{t/2\sqrt{2}}(119\cos2\phi-175\sqrt{2}\sin2\phi)
\nonumber \\
&& \times (3\, \delta r(0)+6\, \delta p_\phi(0)+6\sqrt{2}\, \delta p_r(0))\, ,
\end{eqnarray}
so that
\begin{equation}
\delta \Phi(t) \simeq e^{t/2\sqrt{2}} \delta \Phi(0) \, .
\end{equation}
As promised, the phase difference grows exponentially on a timescale equal to $T_\lambda=2\sqrt{2}$.

\section*{Acknowledgements}
I would like to that Scott Hughes and Janna Levin for several interesting discussions.

\end{document}